\documentclass[aps,twocolumn,a4paper,showpacs]{revtex4}
\usepackage{graphicx}
\usepackage{amsmath}
\usepackage{amssymb}
\usepackage{enumerate}
\usepackage{subfigure}
\usepackage{tabularx}

\newcommand{\be}{\begin{equation}}
\newcommand{\ee}{\end{equation}}
\newcommand{\ben}{\begin{eqnarray}}
\newcommand{\een}{\end{eqnarray}}
\newcommand{\bes}{\begin{subequations}}
\newcommand{\ees}{\end{subequations}}

\newcommand{\bb}{\bibitem}

\newcommand{\LL}{{\mathcal L}}

\newcommand{\vphi}{{\varphi}}
\begin{document}
\title{Split Q-Balls}
\author{D. Bazeia$^{1}$, L. Losano$^{1}$, M.A. Marques$^1$, and R. Menezes$^{2,3}$}
\affiliation{$^1$Departamento de F\'\i sica, Universidade Federal da Para\'\i ba, 58051-970 Jo\~ao Pessoa, PB, Brazil}
\affiliation{$^2$Departamento de F\'\i sica, Universidade Federal de Campina Grande, 58109-970, Campina Grande, PB, Brazil}
\affiliation{$^3$Departamento de Ci\^encias Exatas, Universidade Federal
da Para\'{\i}ba, 58297-000 Rio Tinto, PB, Brazil}

\begin{abstract}
We investigate the presence of non-topological solutions of the Q-ball type in $(1,1)$ spacetime dimensions. The model engenders the global $U(1)$ symmetry and is of the k-field type, since it contains a new term, of the fourth-order power in the derivative of the complex scalar field. It supports analytical solution of the Q-ball type which is stable quantum mechanically. The new solution engenders an interesting behavior, with the charge and energy densities unveiling a splitting profile.
\end{abstract}
\pacs{11.27.+d, 98.80.Cq}
\date{\today}
\maketitle

\section{Introduction}
Defect structures appear in several different contexts of Physics, and can engender topological \cite{b1,b2} or non-topological \cite{tdleefirst,kolb,wilets,tdlee} profile. Usually, in high energy physics the topological defects are stable against small fluctuations around the static solutions, and the non-topological ones are unstable. However, one may sometimes find mechanisms that serve to stabilize the non-topological defects. An important possibility refers to the solution known as Q-ball \cite{coleman}, which represents a charged non-topological configuration where the charge plays central role to its stabilization. 

Over the years, Q-balls have been widely studied \cite{tdlee,coleman,cervero,kk,kusenko1,kusenko2,kusenko3,q1,tuomas,minos,q2,sut,ku,sut1,ed,sta1,sta2,bmm1,bmm2} and the investigations usually require numerical methods, since the equations involved are nonlinear and very hard to be solved. The task is complicated, even in the simplest scenario in which Q-balls appear, in flat spacetime with the scalar field engendering global $U(1)$ symmetry. However, other possibilities are also allowed, and they may be studied, for instance, under the local U(1) symmetry \cite{rosen,gqb,lgqb,gnps} and in curved spacetime \cite{matsuda1,matsuda2,psc}.

An interesting possibility is that Q-balls may contribute to dark matter \cite{kusenko1,kusenko2,silk}, and this 
has inspired us to propose and investigate a new model, with the Lagrange density presenting a non-standard modification in its kinetic term. Although the generalization makes the problem harder, we show that it supports Q-ball solutions which can be found analytically. Another motivation for the change considered in this work is suggested by Cosmology, where the so-called k-field modification of the kinematics of the scalar field appeared before in, e.g., \cite{kf1,kf2,kf3,kf4,bab,baz,ada}, as a way to respond to the accelerated cosmic expansion detected in \cite{acce1,acce2}. As we are going to show, the proposed model presents new features, different from the ones that appear in the standard model, and we find non-topological solution of the Q-ball type, which is identified as a split Q-ball. For this reason, in order to better expose the difference between a Q-ball and a
split Q-ball, in Sec.~\ref{sm} we start reviewing the basic concepts and properties of the standard Q-ball, and then, in Sec.~\ref{newscenario} we go on and introduce the new model and study the main features it comprises, emphasizing the differences concerning the standard model. We end the work with comments and conclusions in Sec.~\ref{conclusions}. 

\section{Standard Model}\label{sm}

To investigate global Q-balls, we consider the Lagrange density for a complex scalar field in $(1,D)$ Minkowski spacetime
\be
{\cal L} = \frac12 \partial_\mu{\vphi^*} \partial^\mu \vphi - V(|\vphi|),
\ee
where $V(|\vphi|)$ is the potential. The associated equation of motion is given by
\be\label{seomd}
\ddot{\vphi} - \nabla^2 \vphi + \frac{\vphi}{|\vphi|} \frac{dV}{d|\vphi|} = 0,
\ee
where dot stands for time derivative, as usual. The conserved current and energy-momentum tensor are given by
\bes\label{sjdd}\ben\label{sjd}
j_\mu &=& \Im \left(\vphi^*\partial_\mu\vphi\right), \\
\label{std}
T_{\mu\nu} &=& \Re\left(\partial_\mu{\vphi^*}\partial_\nu\vphi\right)  - \eta_{\mu\nu}{\mathcal L},
\een\ees
where $\Re(z)$ and $\Im(z)$ represents the real and imaginary parts of $z$, respectively. We now consider the Q-ball ansatz in $(1,1)$ spacetime dimensions, which reads
\be\label{ansatz}
\vphi(x,t)=\sigma(x)\,e^{i\omega t}.
\ee
The equation of motion \eqref{seomd} becomes
\be\label{seom}
\sigma^{\prime\prime} =\frac{dV}{d\sigma} -\omega^2\sigma,
\ee
with the boundary conditions
\ben\label{bcond}
\sigma^\prime(0) = 0;\;\;\;\;\;\sigma(\infty) = 0,
\een
which one uses to search for non-topological solutions. We can integrate Eq.~\eqref{seom} to get the effective equation
\be\label{seffeq}
\frac12{\sigma^{\prime}}^2 = U,
\ee 
with $U=U(\sigma)$ being an effective potential for the field $\sigma$.  It has the form
\be\label{sveff}
U(\sigma) = V (\sigma)- \frac12\omega^2\sigma^2.
\ee
We search for solutions of the effective equation that connects a minimum, which we consider at $\sigma=0$, and a zero of $U(\sigma)$. Note that $\omega$ plays an important role in this game. The boundary conditions \eqref{bcond} are obeyed if $\omega$ is in the interval
\be\label{scondomega}
\omega_-<\omega<\omega_+,
\ee
with $\omega_+ = V^{\prime\prime}(0)$ and $\omega_-=\sqrt{2V(\sigma_0)/\sigma_0^2}$, where $\sigma_0$ is the minimum of $V(\sigma)/\sigma^2$.

The components of the current \eqref{sjd} are
\be\label{schargedens}
j_0 = \omega\sigma^2 \quad \text{and} \quad j_1 = 0.
\ee
It is straightfoward to show that the charge is conserved and is given by
\be\label{scharge}
Q= \omega \int_\infty^\infty{dx\, \sigma^2}.
\ee

By using the ansatz \eqref{ansatz}, the non-vanishing components of the energy-momentum tensor can be cast to the form
\be
T_{00} = \epsilon_k + \epsilon_g + \epsilon_p, \quad \text{and} \quad T_{11} =  \epsilon_k + \epsilon_g - \epsilon_p,
\ee
where we are using 
\be
\epsilon_k = \frac{1}{2}\omega^2\sigma^2, \quad \epsilon_g=\frac12{\sigma^\prime}^2, \quad \epsilon_p = V(\sigma),
\ee
to represent the kinetic, gradient and potential portions of the energy density, respectively. The kinetic energy density can be written in terms of the charge density, $j_0$, from Eq.~\eqref{schargedens}, as $\epsilon_k = \omega j_0/2$. This makes the energy to have an explicit dependence on the charge \eqref{scharge}. The energy-momentum tensor is conserved, i.e., $\partial_\mu T^{\mu\nu}=0$, which makes $T_{11}$ constant. Furthermore, localized stressless solutions are obtained with $T_{11}=0$., and this gives $\epsilon_p=\epsilon_k + \epsilon_g$, so the energy density can be written as $T_{00} = 2(\epsilon_k + \epsilon_g)$. Furthermore, we can understand the energy density in the form
\be\label{t00q}
T_{00} = \epsilon_Q + \epsilon_I,
\ee
where
\be
\epsilon_Q = \omega j_0 \quad \text{and} \quad \epsilon_I = {\sigma^\prime}^2,
\ee
with $\epsilon_Q$ and $\epsilon_I$ being the two contributions to the energy density, the first being explicitly dependent on the charge density, and the second independent of the charge density.

By taking the potential
\be\label{pot}
V(|\vphi|)=  \frac12 |\vphi|^2 - \frac13|\vphi|^3 + \frac14 a\,|\vphi|^4,
\ee
where $a$ is a real parameter, the effective potential is given by
\be\label{veff1}
U(\sigma) = \frac12 (1-\omega^2)\sigma^2 -\frac13 \sigma^3 + \frac14 a\,\sigma^4.
\ee
This model was studied in \cite{cervero,kk} and the solution can be written in the form \cite{bmm1}
\ben\label{ssolst}
\sigma(x) &=& \sqrt{\frac{1-\omega^2}{2a}}\left[\tanh\left(\frac12 \sqrt{1-\omega^2}\, x + b \right)+ \nonumber\right. \\
&&\left.-\tanh\left(\frac12 \sqrt{1-\omega^2}\, x - b \right)\right],
\een
where
\be
b=\frac12 \text{arctanh}\left({3\sqrt{(1-\omega^2)a/2}}\right).
\ee
The expression in Eq.~\eqref{ssolst} is exact solution of Eq.~\eqref{seom} that obeys the boundary conditions \eqref{bcond}. In this case, Eq.~\eqref{scondomega} holds with $\omega_-=\sqrt{1-2/(9a)}$ and $\omega_+=1$. By using the exact solution \eqref{ssolst}, it is possible to calculate the charge from Eq.~\eqref{scharge}; the result can be written as
\be\label{qst}
Q=\frac{4\omega\sqrt{1-\omega^2}}{a} \left(2b\coth(2b)-1\right).
\ee
From the above expression, one sees that $Q\to \infty$ if $\omega\to\omega_-$, except for $a=2/9$, which gives $Q\to0$ for $\omega\to\omega_-$. For $\omega\to\omega_+$, $Q\to0$, for any $a$. We can calculate the total energy integrating Eq.~\eqref{t00q}, with $E_Q=\omega Q$ and using
\ben
E_I\!\!&=&\!\! \frac{2(1-\omega^2)^{3/2}}{3a} \nonumber \\
\!\!& \times &\!\! \left(\frac{1+3(8b+3)e^{4b} + 3(8b-3)e^{8b} - e^{12b}}{\left(1 - e^{4b}\right)^3}\right)\!.\;\;
\een
The result is $E=E_Q+E_I$. The full expression shows that the criteria for quantum mechanical stability, $E/Q<\omega_+$, is not fulfilled for $a_0\equiv 2/9\approx0.22222$, only for $a>a_{s1}\equiv 0.22268$. For $a=a_{s2}\equiv 0.22540$ the local maximum of the charge becomes an inflection point, and for $a>a_{s2}$, it is a monotonically decreasing function. 

\section{New Model}\label{newscenario}
Let us now focus on the generalized model, which is taken as the Lagrange density
\be
\LL = \frac12\partial_\mu\vphi^*\partial^\mu\vphi - \frac14{f(|\vphi|)} \left(\partial_\mu\vphi^*\partial^\mu\vphi\right)^2- V(|\vphi|),
\ee
where $f(|\vphi|)$ is in principle an arbitrary function of the complex field; if it vanishes, we get back to the standard case. The model is inspired by Refs.~\cite{gen1,gen2}, and besides the standard kinematical term, it contains a modification that makes it much harder to be investigated analytically. This comes from the presence of the function $f(|\varphi|)$ and from the other factor, that adds higher-order power in the derivative of the scalar field. This kinematical modification appears as a k-field contribution, so we search for generalized non-topological solutions of the $Q$-ball type. 

To study the model, one varies its action with respect to the field, to obtain the equation of motion
\ben\label{eomf}
&&\partial_\mu\left(\left(1-f\partial_\nu\vphi^*\partial^\nu\vphi\right) \partial^\mu\vphi\right)+ \nonumber \\ \label{geom}
&&+ \frac{\vphi}{|\vphi|} \left(V_{|\vphi|} +  \frac{f_{|\vphi|}}{4} \left(\partial_\mu\vphi^*\partial^\mu\vphi\right)^2 \right)=0.
\een
The conserved current and energy-momentum tensor are given by
\bes\ben\label{gjd}
j_\mu &=& \left(1-f\partial_\nu\vphi^*\partial^\nu\vphi\right) \Im \left(\vphi^*\partial_\mu\vphi\right), \\
\label{gtd}
T_{\mu\nu} &=& \left(1-f\partial_\alpha\vphi^*\partial^\alpha\vphi\right)\Re\left(\partial_\mu{\vphi^*}\partial_\nu\vphi\right) - \eta_{\mu\nu}{\mathcal L},\;\;\;
\een\ees
where $\Re(z)$ and $\Im(z)$ represents the real and imaginary parts of $z$, as in \eqref{sjdd}.

With the ansatz \eqref{ansatz}, the equation of motion \eqref{geom} becomes
\ben\label{geomansatz}
\left(1\!-\!(\omega^2\sigma^2\!-\!3{\sigma^\prime}^2)f\right)\!\sigma^{\prime\prime}\!+\!\left({\sigma^\prime}^2f_\sigma\! \!- \omega^2\sigma(\sigma f_\sigma\!+\!2f)\right) {\sigma^\prime}^2\!\!+\nonumber \\
 +\, \omega^2\sigma \left(1\!-\!(\omega^2\sigma^2\!-\!{\sigma^\prime}^2)f\right) =V_\sigma + \frac{f_\sigma}{4} (\omega^2\sigma^2\!-\! {\sigma^\prime}^2)^2.\;\;\;
\een
One notes that it is much more involved than in the standard case, which appears in the limit $f\to0$,
given by \eqref{seom}.

From the temporal components of Eqs.~\eqref{gjd} and \eqref{gtd}, one gets the following charge and energy densities, respectively,
\bes\ben\label{gj0ansatz}
j_0 &\!=\!& \omega\sigma^2-f\omega\sigma^2(\omega^2\sigma^2-{\sigma^\prime}^2),\\ \label{gt00ansatz}
T_{00} &\!=\!& \frac12\omega^2\sigma^2 + \frac12{\sigma^\prime}^2+V +\nonumber\\
&&-\frac{f}{4}(3\omega^2\sigma^2+{\sigma^\prime}^2)(\omega^2\sigma^2-{\sigma^\prime}^2).
\een\ees
It is possible to integrate Eq.~\eqref{geomansatz} to get
\be
\frac{3f}{4}{\sigma^\prime}^4 + \frac{1-f\omega^2\sigma^2}{2}{\sigma^\prime}^2 = V-\frac12\omega^2\sigma^2+\frac14 f\omega^4\sigma^4.
\ee
This equation is hard to solve, and we have tried different possibilities, with no success. However, we noted that if one chooses $f=1/(4V)$, it gives two possibilities for ${\sigma^\prime}^2$, one of them providing an interesting solution, which is described by the effective first-order equation
\be\label{geffeq}
\frac12 {\sigma^\prime}^2 = U(\sigma)
\ee
where
\be
U(\sigma) = \frac23 V(\sigma)-\frac16\omega^2\sigma^2.
\ee
Surprisingly, one notes that the scalings $x\to x\sqrt{2/3}$ and $\omega\to\omega/\sqrt{2}$ change the above Eq.~\eqref{geffeq} back to Eq.~\eqref{seffeq}. By using Eq.~\eqref{geffeq}, the charge and energy densities \eqref{gj0ansatz} and \eqref{gt00ansatz} become
\bes\ben\label{gj0s}
j_0 &=& \frac{4\omega{\sigma^\prime}^2\sigma^2}{\omega^2\sigma^2+3{\sigma^\prime}^2}, \\\label{gt00s}
T_{00} &=& \frac{4{\sigma^\prime}^2(\omega^2\sigma^2+{\sigma^\prime}^2)}{\omega^2\sigma^2+3{\sigma^\prime}^2}.
\een\ees
Again, we can write the energy density as in Eq.~\eqref{t00q}. Here, we still have $\epsilon_Q = \omega j_0$. However, the other contribution changes to
\be\label{epsilonI}
\epsilon_I = \frac{4{\sigma^\prime}^4}{\omega^2\sigma^2+3{\sigma^\prime}^2}.
\ee
By using the first-order equation \eqref{geffeq}, one can show that the charge and the charge independent contribution to the energy can be written as
\ben\label{qsigma}
Q &=& \frac{2\omega}{3}\int_0^A\frac{\sigma^2\sqrt{12V-3\omega^2\sigma^2}}{V}d\sigma, \\\label{eisigma}
E_I &=& \frac{2}{27}\int_0^A\frac{(12V-3\omega^2\sigma^2)^{3/2}}{V} d\sigma,
\een
where $A$ is the amplitude of the solution, calculated through the equation $V(A)=\omega^2A^2/4$.

\begin{figure}[htb!]
\includegraphics[width=4.2cm]{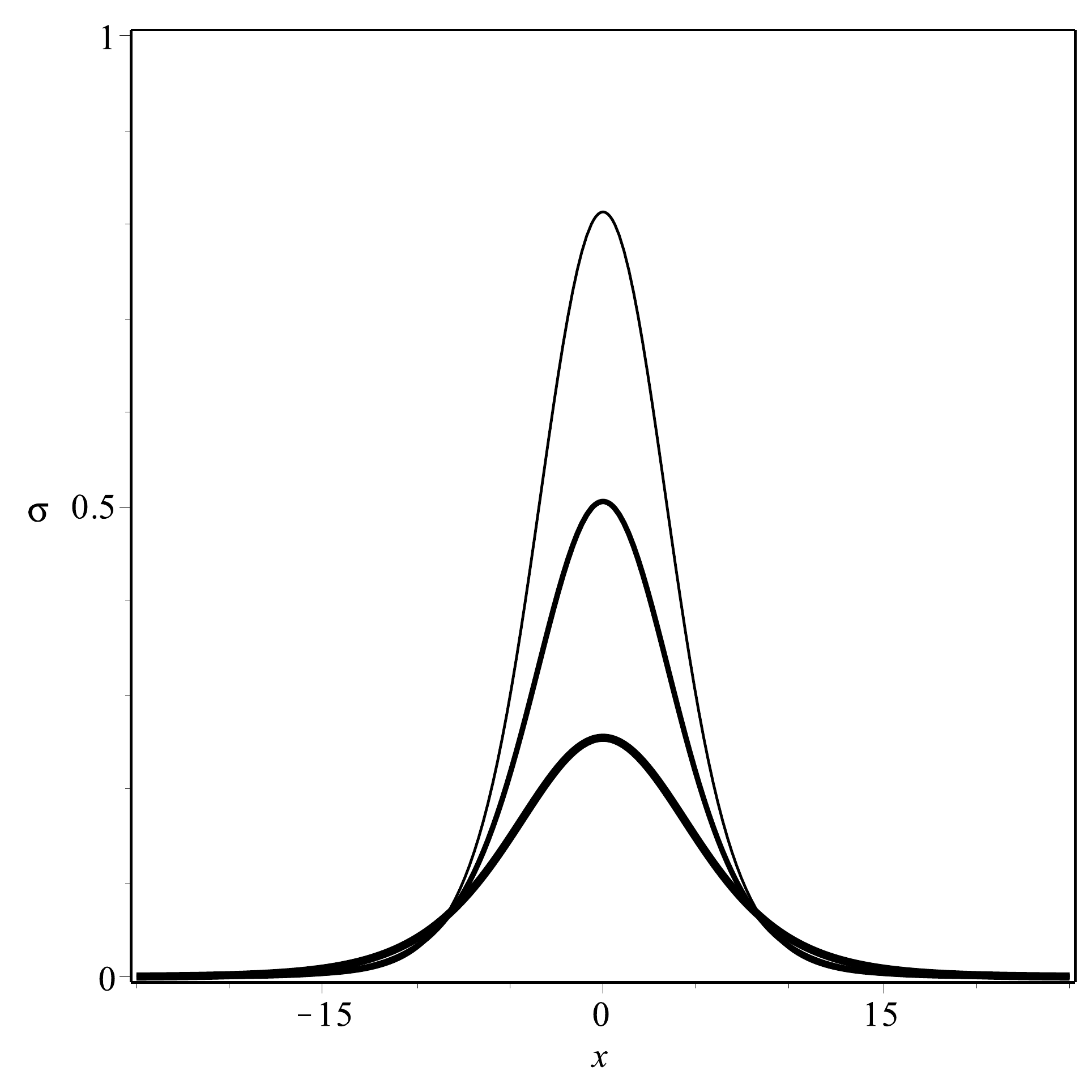}
\includegraphics[width=4.2cm]{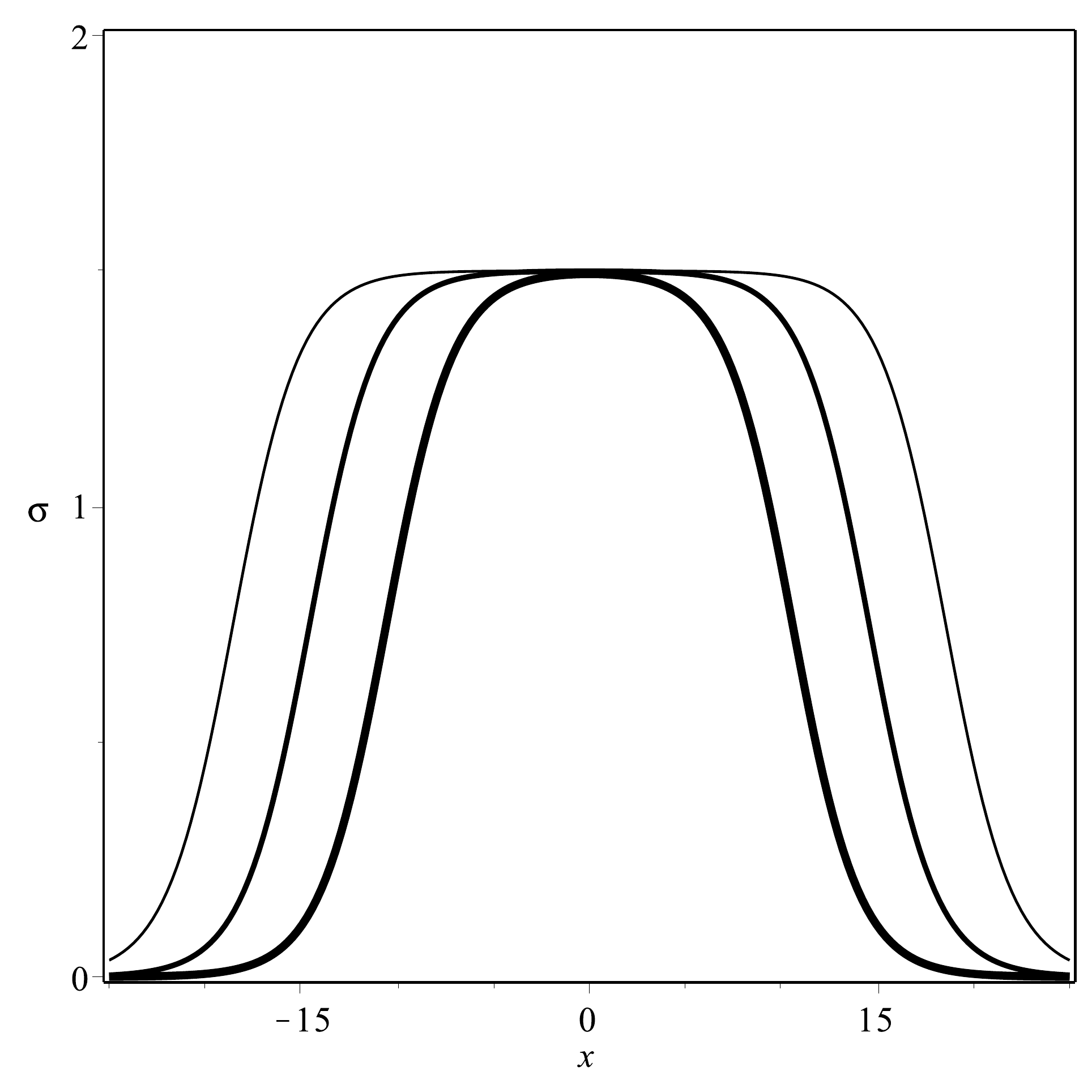}
\caption{The solution \eqref{ssol} depicted for $a=4/9$, $\omega=1.1,1.2$ and $1.3$ (left) and $\omega=1 + \delta$, with $\delta=10^{-9},10^{-7}$ and $10^{-5}$ (right). In each plot, the thickness of the lines increases with $\omega$.}\label{figssol}
\end{figure}

Considering the potential \eqref{pot}, the solution in this case is given by
\ben\label{ssol}
\sigma(x) &=& \frac12\sqrt{\frac{2-\omega^2}{a}}\left[\tanh\left(\frac12 \sqrt{\frac{2-\omega^2}{3}}\, x + b \right)+ \nonumber\right. \\
&&\left.-\tanh\left(\frac12 \sqrt{\frac{2-\omega^2}{3}} x - b \right)\right],
\een
where
\be
b=\frac12 \text{arctanh}\left({\frac32\sqrt{(2-\omega^2)a}}\right).
\ee
It is depicted in Fig.~\ref{figssol}, and one notes that $\omega$ is bounded by Eq.~\eqref{scondomega} with
$\omega_-=\sqrt{2-4/(9a)}$ and $\omega_+=\sqrt{2}$. In this case, the amplitude of the solution is given by
\be\label{amplitude}
A=\frac{2-\sqrt{4-9a(2-\omega^2)}}{3a}.
\ee
When $\omega\to\omega_-$, it assumes the maximum value, given by $2/(3a)$. As one increases $\omega$, it gets smaller and smaller, going down to zero for $\omega$ approaching $\omega_+$. Thus, the amplitude of the solution is limited. In Fig.~\ref{figssol} we plot the solution \eqref{ssol} for $a=4/9$ and for several values of $\omega$. We see that a plateau of height $2/(3a)$ appears as $\omega\to\omega_-$. This limit is equivalent to the thin wall limit in Ref.~\cite{coleman}. As $\omega$ approaches $\omega_-$, the plateau gets wider.

\begin{figure}[t!]
\includegraphics[width=4.2cm]{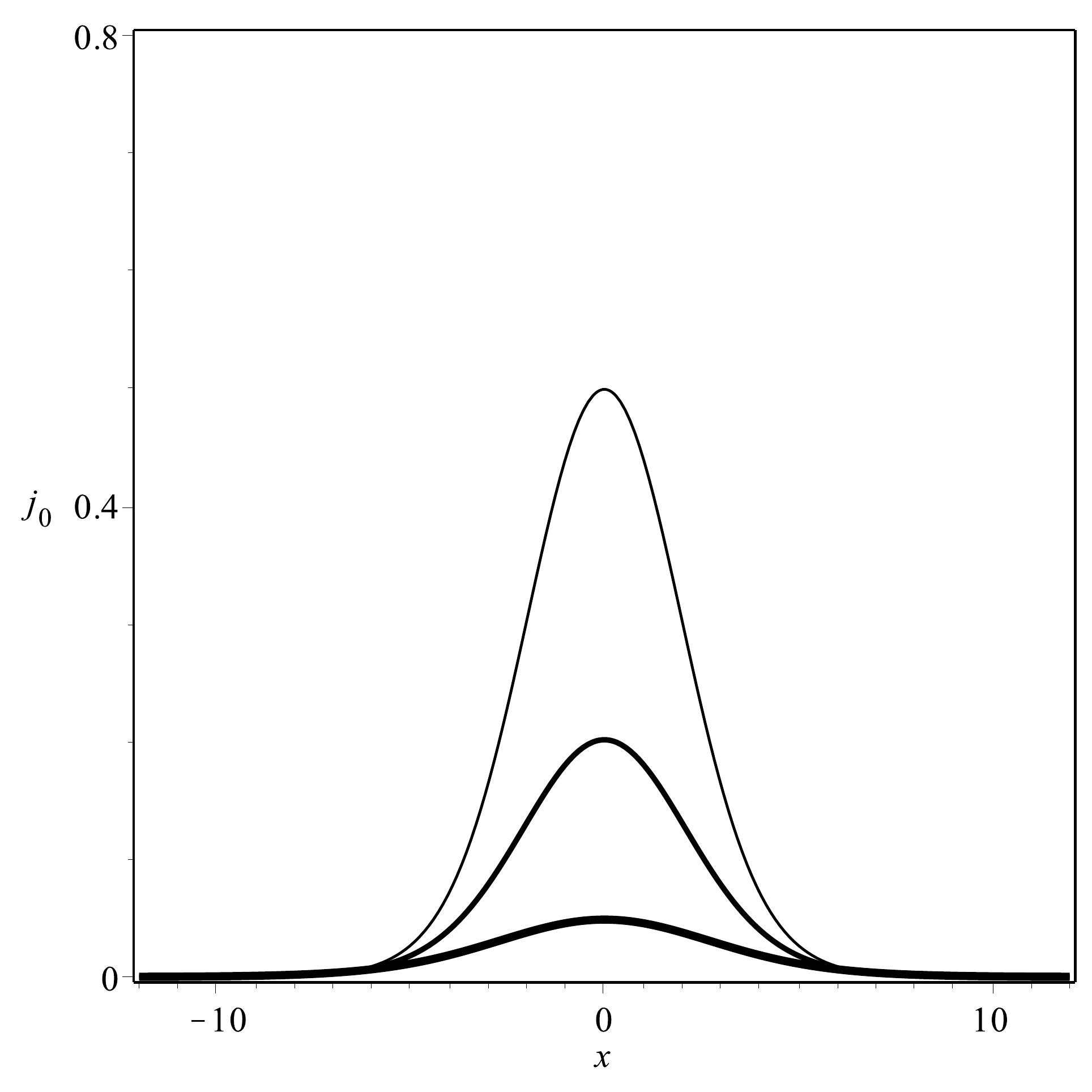}
\includegraphics[width=4.2cm]{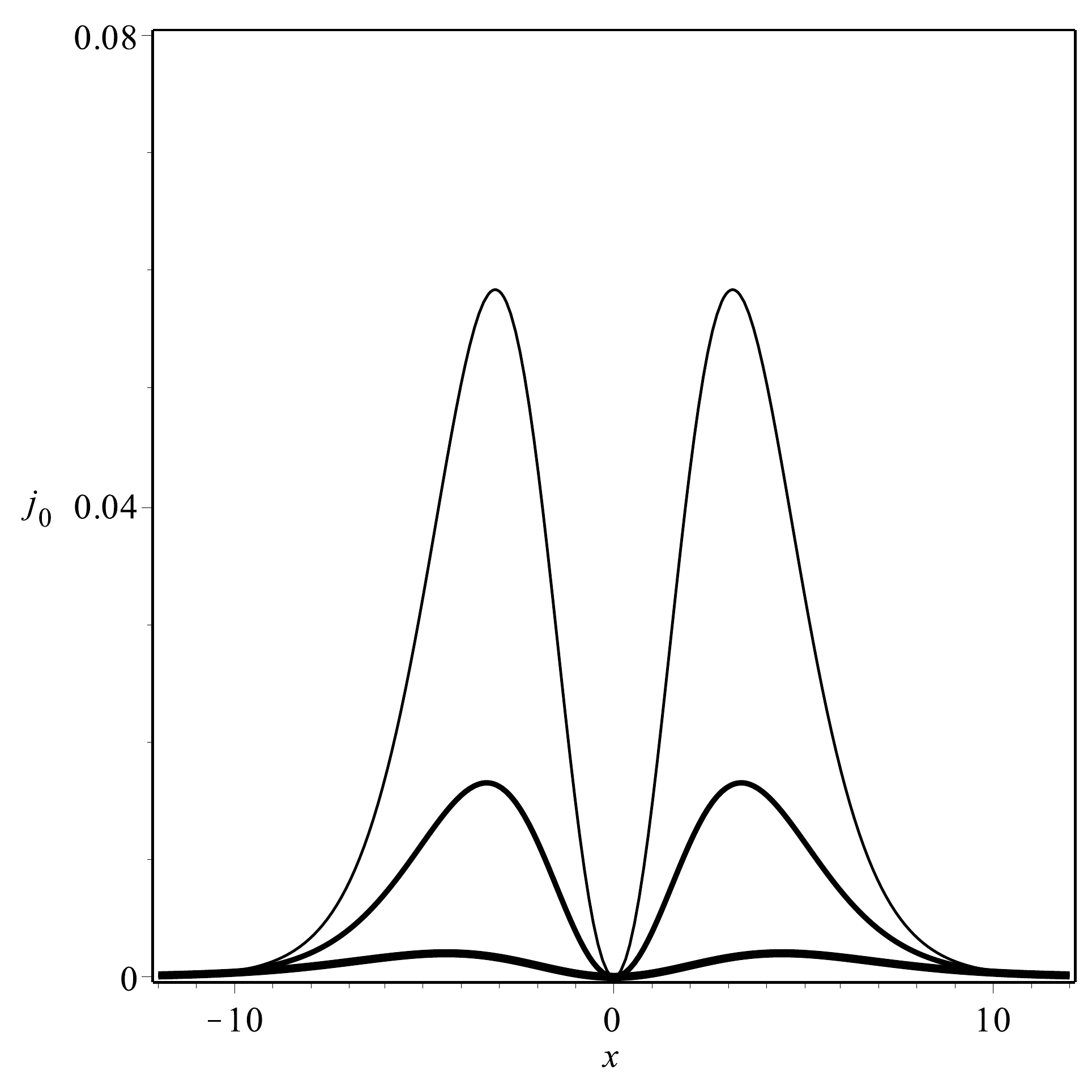}
\includegraphics[width=4.2cm]{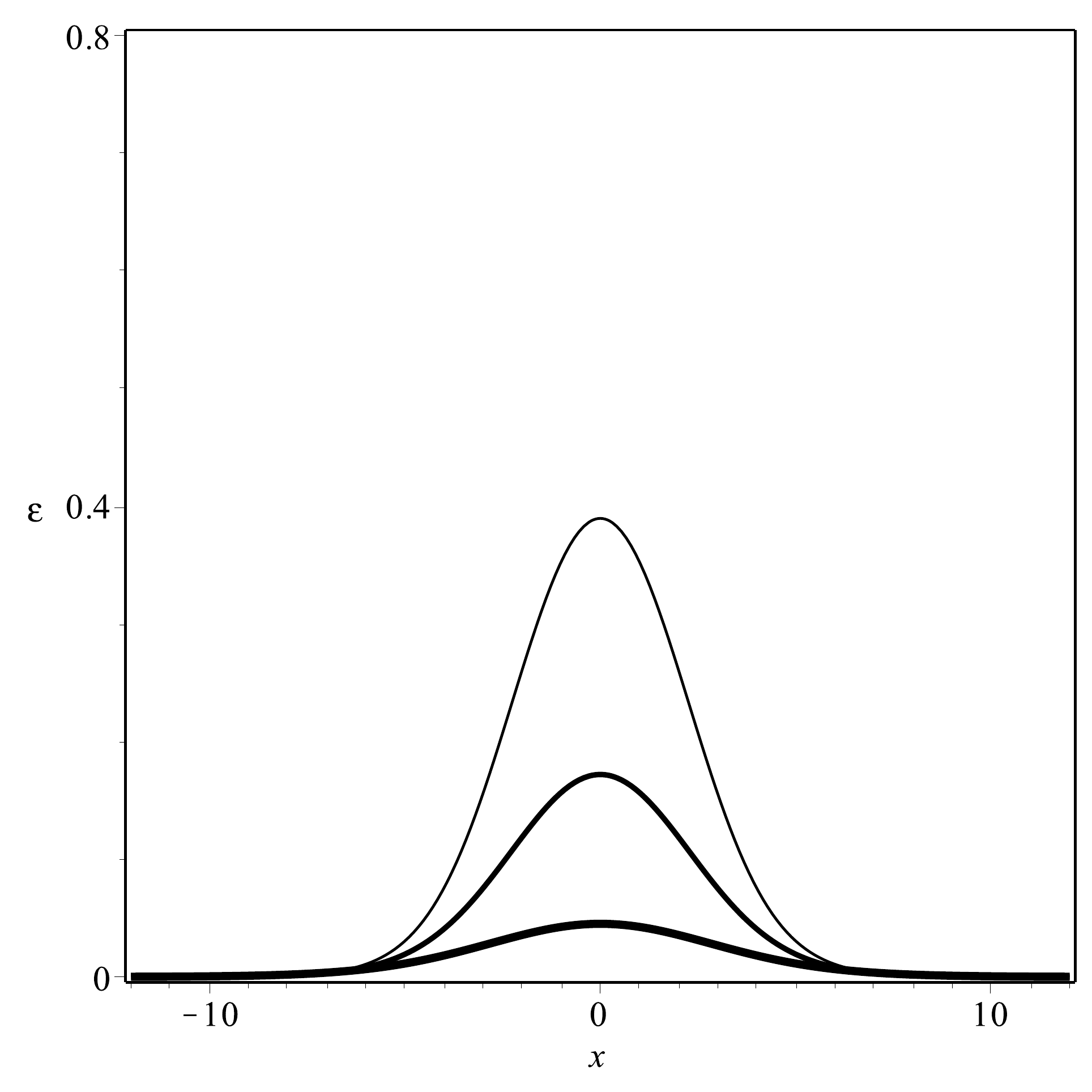}
\includegraphics[width=4.2cm]{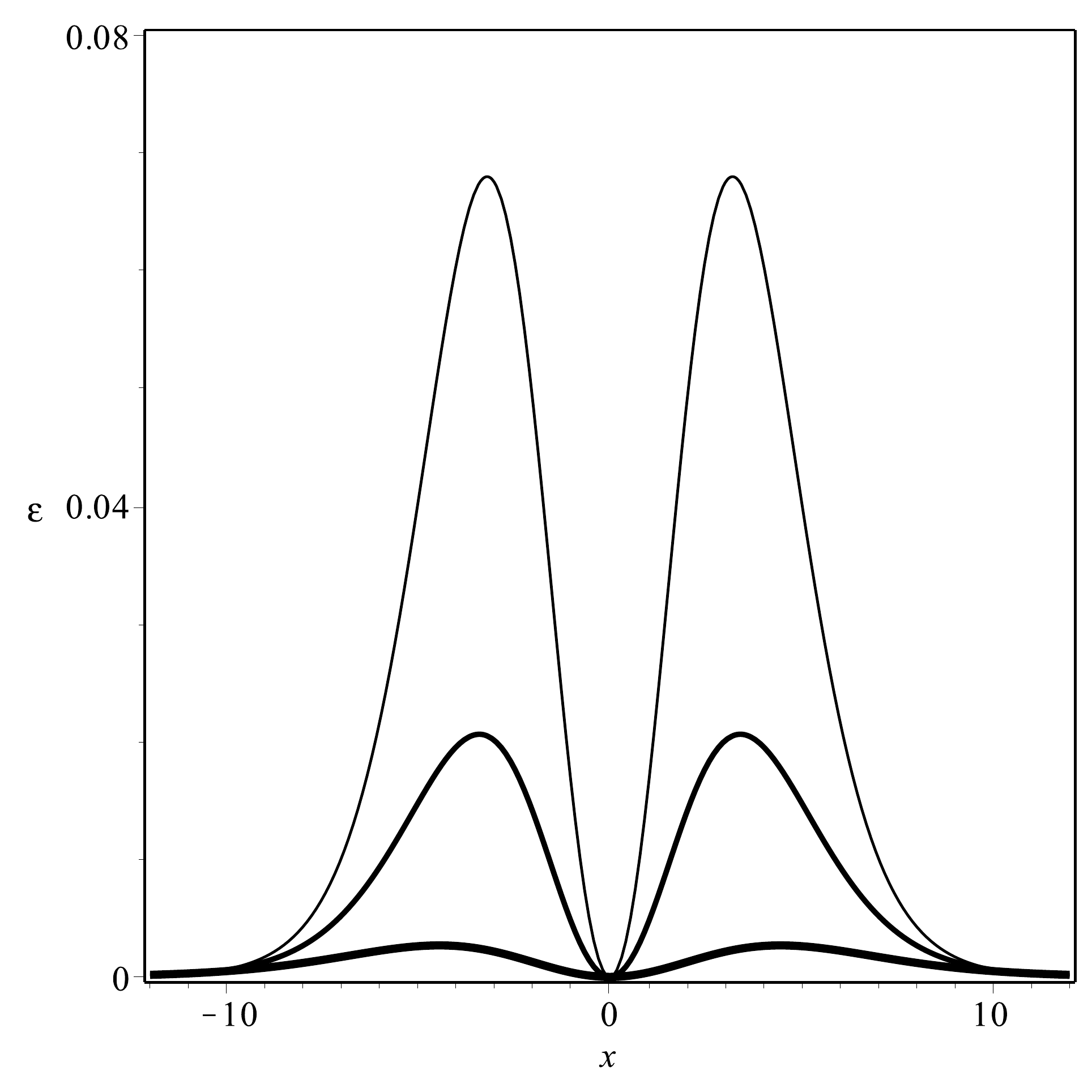}
\caption{The charge (top) and energy (bottom) densities in the standard (left) and generalized (right) models, depicted for $a=4/9$ and $\omega=\omega_- + (\omega_+-\omega_-)n/4$, with $n=1,2$ and $3$. In each plot, the thickness of the line increases with $\omega$.}
\label{figdensities}
\end{figure}

We use Eqs.~\eqref{gj0s} and \eqref{gt00s} to see that the charge and energy densities split, as depicted in Fig.~\ref{figdensities}. This is a difference to be highlighted in the generalized Q-ball. Although the solution \eqref{ssol} is essentially the standard solution \eqref{ssolst}, the charge and energy densities acquire distinct new profiles: while in the standard case they are bell-shaped, in the generalized model they appear to split into two portions around its center, controlled by $\omega$. Although the tendency to split was also identified before in \cite{bmm1}, in a model with standard kinematics, here the effect is stronger and simulates an internal structure. A similar behavior appeared in the braneworld model investigated in \cite{brane}, with gravity modified to accommodate the change $R\to R+\alpha R^2$, with $R$ being the scalar curvature and $\alpha$ a real parameter used to control deviation from the standard situation. 

\begin{figure}[t!]
\includegraphics[width=6cm]{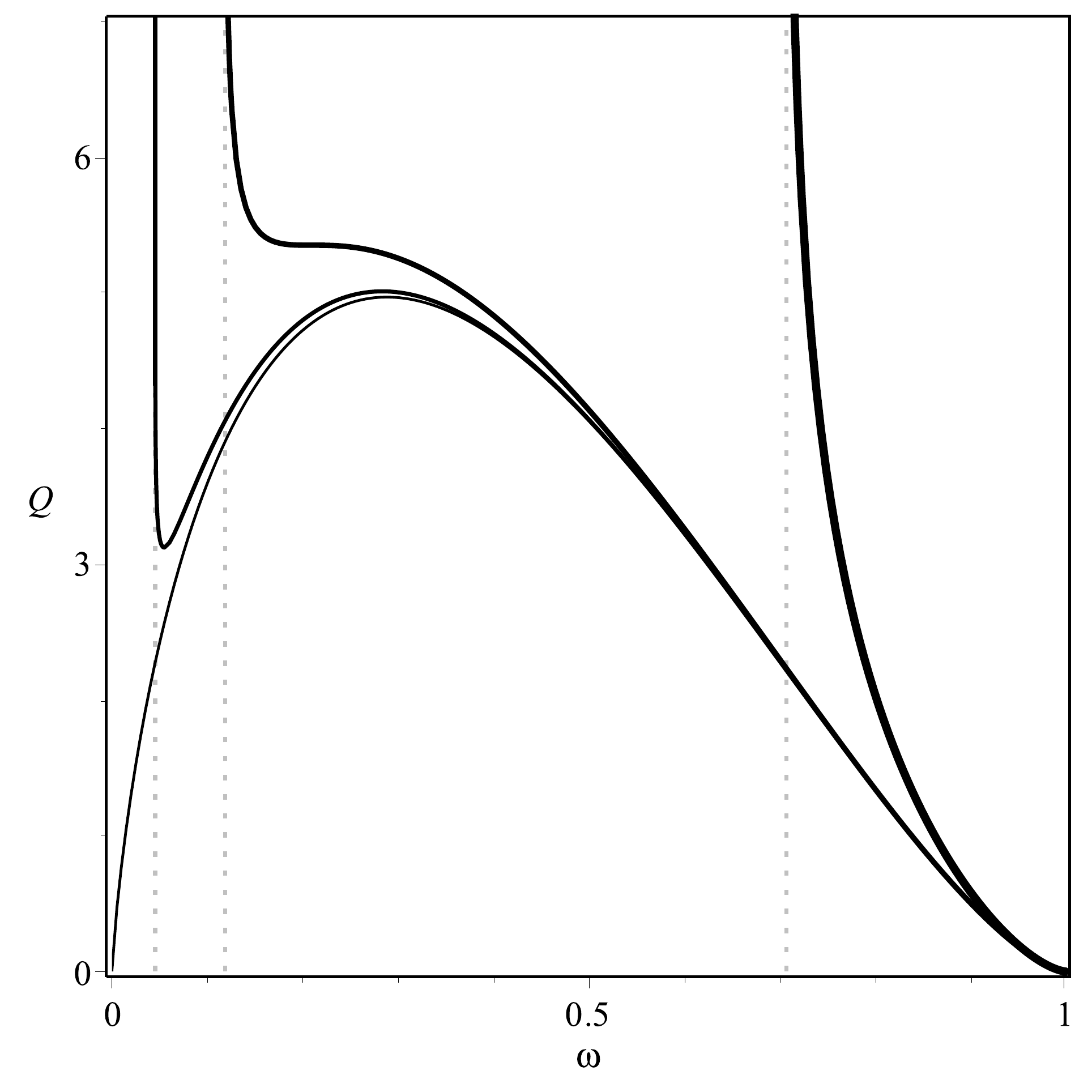}
\includegraphics[width=6cm]{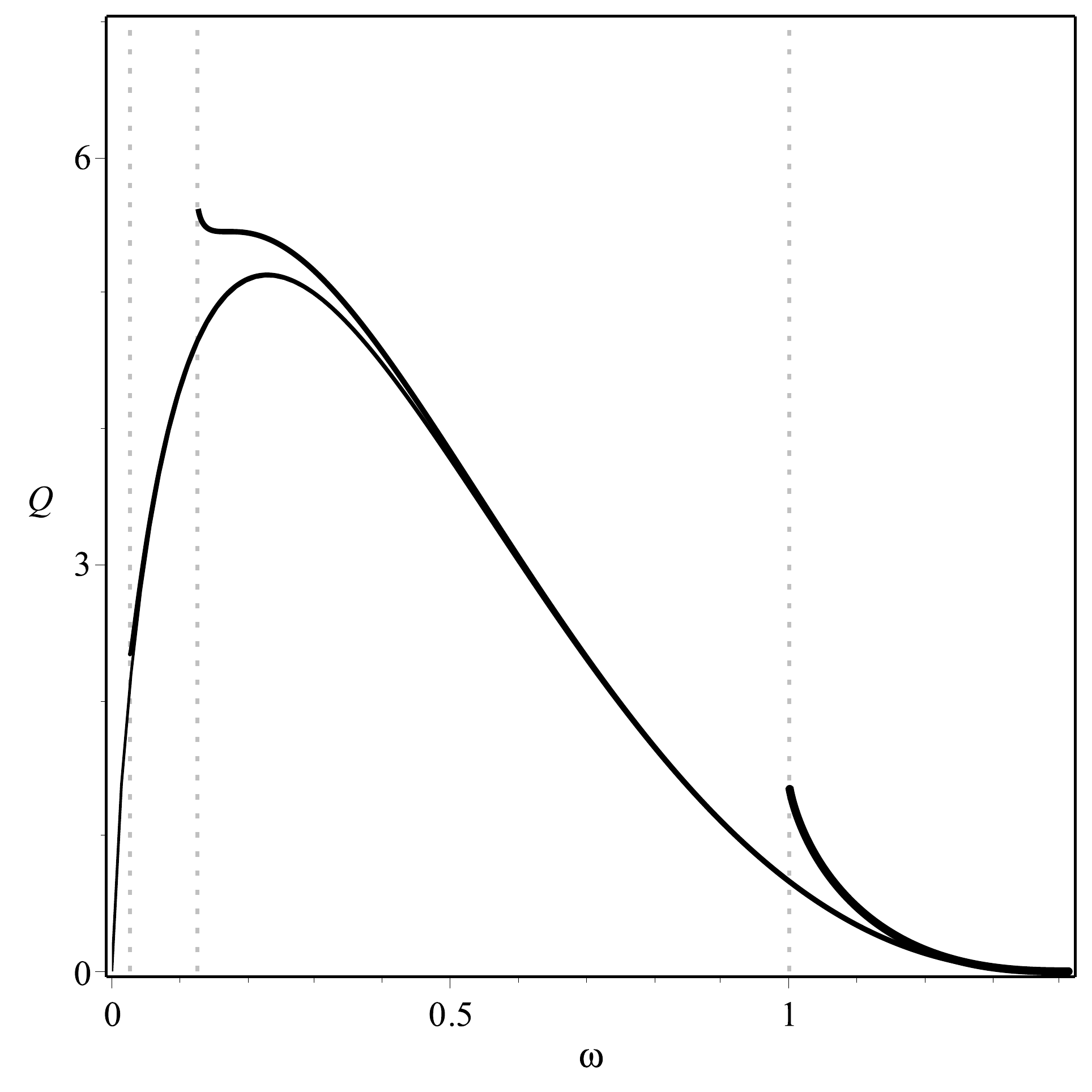}
\caption{The charge in the standard model (top) for $a=a_0,a_{s1},a_{s2}$ and $a_3$, and in the generalized model (bottom) for $a=a_0,a_1,a_2$ and $a_3$. In each plot, the thickness of the lines increases with $a$.}\label{cplot}
\end{figure}

In the generalized model in consideration, the Eqs.~\eqref{qsigma} and \eqref{eisigma} allow to calculate analytical expressions for the charge and the charge independent portion of the energy. However, they are cumbersome for $a$ arbitrary, so we illustrate the general situation describing the expressions in the case $a=4/9$; they have the form
\ben\label{chargean}
Q \!\! &=& \!\! 6\sqrt{3} \omega\left(\frac12\ln(1-\Omega^2\omega^2) - \ln(1-\Omega\, \omega)-
\Omega\, \omega \right)+ \nonumber\\
\!\! && +  6\sqrt{3} \omega^2\left(\text{arctan}\left(\Omega\right)-\text{arctanh}\left(\Omega \right) \right),
\een
and
\ben
E_I\!\! &=& \!\! \frac{\sqrt{3}}{6}\left(3\ln(1-\Omega^2\omega^2) - 6\ln(1-\Omega\,\omega) -2\Omega\,\omega \right)+ \nonumber\\
 &&-\frac{\sqrt{3}\omega^2}{6}\left(9\ln(1\!-\Omega^2\omega^2)\!-\!18\ln(1-\Omega\omega)\!-\! 16\Omega\omega \right)\!+ \nonumber \\
&&-2\sqrt{3}\,\omega^3\!\left(\text{arctan}\left(\Omega\right)\!-
\!\text{arctanh}\left(\Omega \right) \right).\!\!\!
\een
In the above expressions, we have set $\Omega=\sqrt{2-\omega^2}/\omega$. When $\omega\to\omega_+$, we see that $Q\to0$ and $E_I\to0$. However, when $\omega\to\omega_-$, $Q\to\sqrt{3} (3\pi/2 + 3\ln(2) - 6) \approx 1.37149$ and $E_I \to \sqrt{3}(7/3 -\pi/2-\ln(2)) \approx 0.12019$. In Fig.~\ref{cplot} we plot $Q$ for $a=a_0,a_1\equiv 0.22230, a_2\equiv0.22400$ and $a_3\equiv4/9\approx0.44444$. Also, we compare it to the standard case, given by Eq.~\eqref{qst}. We see from Fig.~\ref{cplot} that, for $a>2/9$, the standard case diverges as $\omega$ approaches  $\omega_-$, while the generalized model gives a finite result. In the other cases, the two charges behave similarly, going to zero as $\omega$ approaches $\omega_+$. 
\begin{figure}[htb!]
\includegraphics[width=6cm]{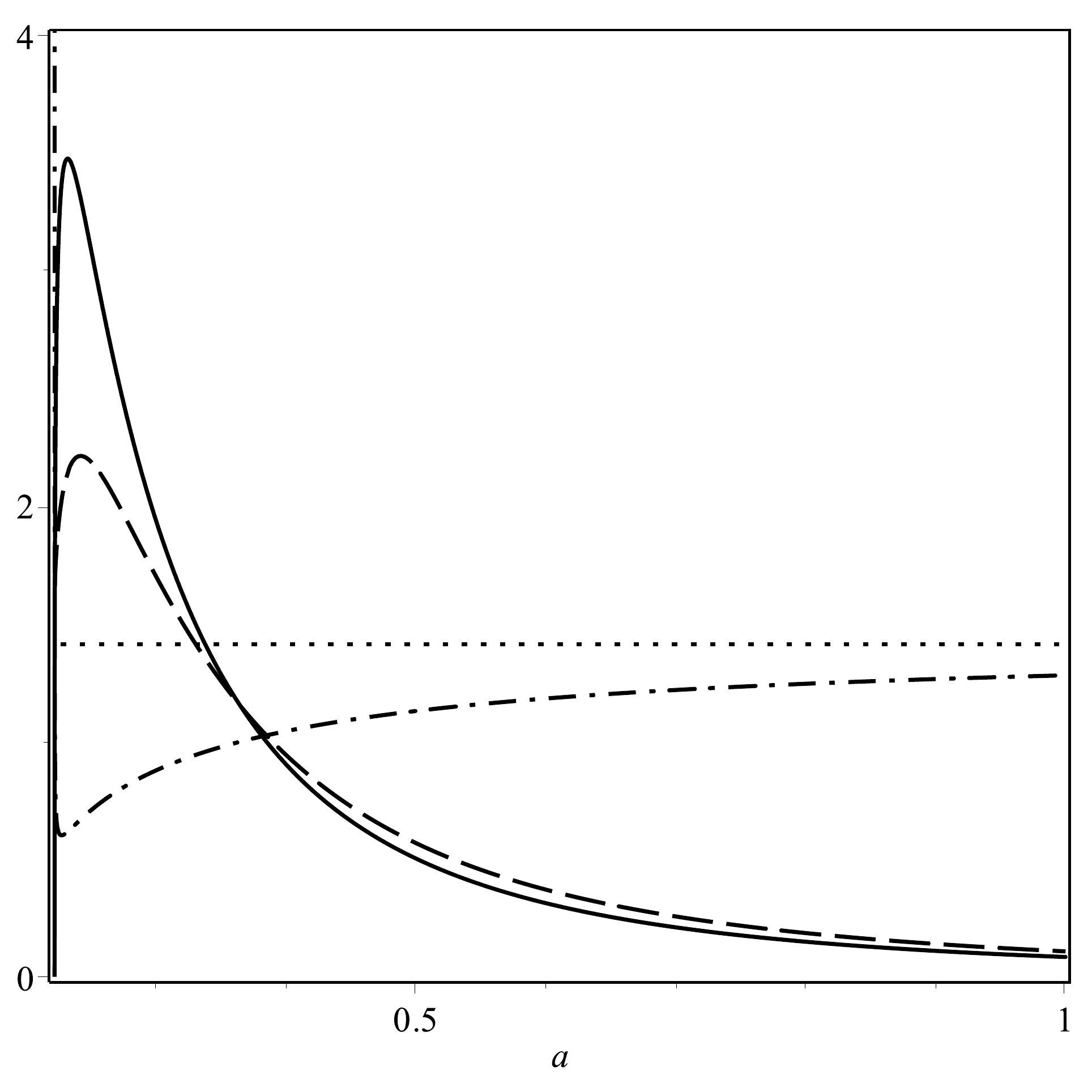}
\caption{The charge, energy and the ratio $E/Q$ in the modified model, calculated in the limit $\omega\to\omega_-$ as a function of $a$, depicted with solid, dashed and dot-dashed lines, respectively. The dotted line represents
$\omega_+$.}\label{cmplot}
\end{figure}

We were able to calculate the charge and the charge-independent energy in the limit $\omega\to\omega_-$ for any $a$. They are given by
\ben\label{qwm}
Q_{\omega_-} &=& \frac{4\sqrt{3}}{27a^2} \left(\sqrt{18a-4}\left(\ln\left(\frac{9a}{9a-2}\right)-2\right)   \right. +\nonumber \\
&&\left.+\,  (18a-4) \arctan\!\left(\frac{2}{\sqrt{18a-4}}\right)\right), \label{eiwm}
\een
and
\ben
E_{I\omega_-} &=& \frac{8\sqrt{3}}{729a^{5/2}}\left(54a-10-3(9a-2)\ln\left(\frac{9a}{9a-2}\right) \right. +\nonumber\\
&& \left.+\frac{6(9a-2)^2}{\sqrt{18a-4}}\arctan\!\left(\frac{2}{\sqrt{18a-4}}\right)\right).
\een
This also shows that the energy is finite if $\omega\to\omega_-$. In Fig.~\ref{cmplot} we plot the charge, energy and the ratio $E/Q$ calculated at $\omega_-$, as a function of $a$. We see that, even though the charge and energy are finite, the ratio $E/Q$ diverges for $a=2/9$, a fact that affects the stability of the solution.

\begin{figure}[htb!]
\includegraphics[width=4.2cm]{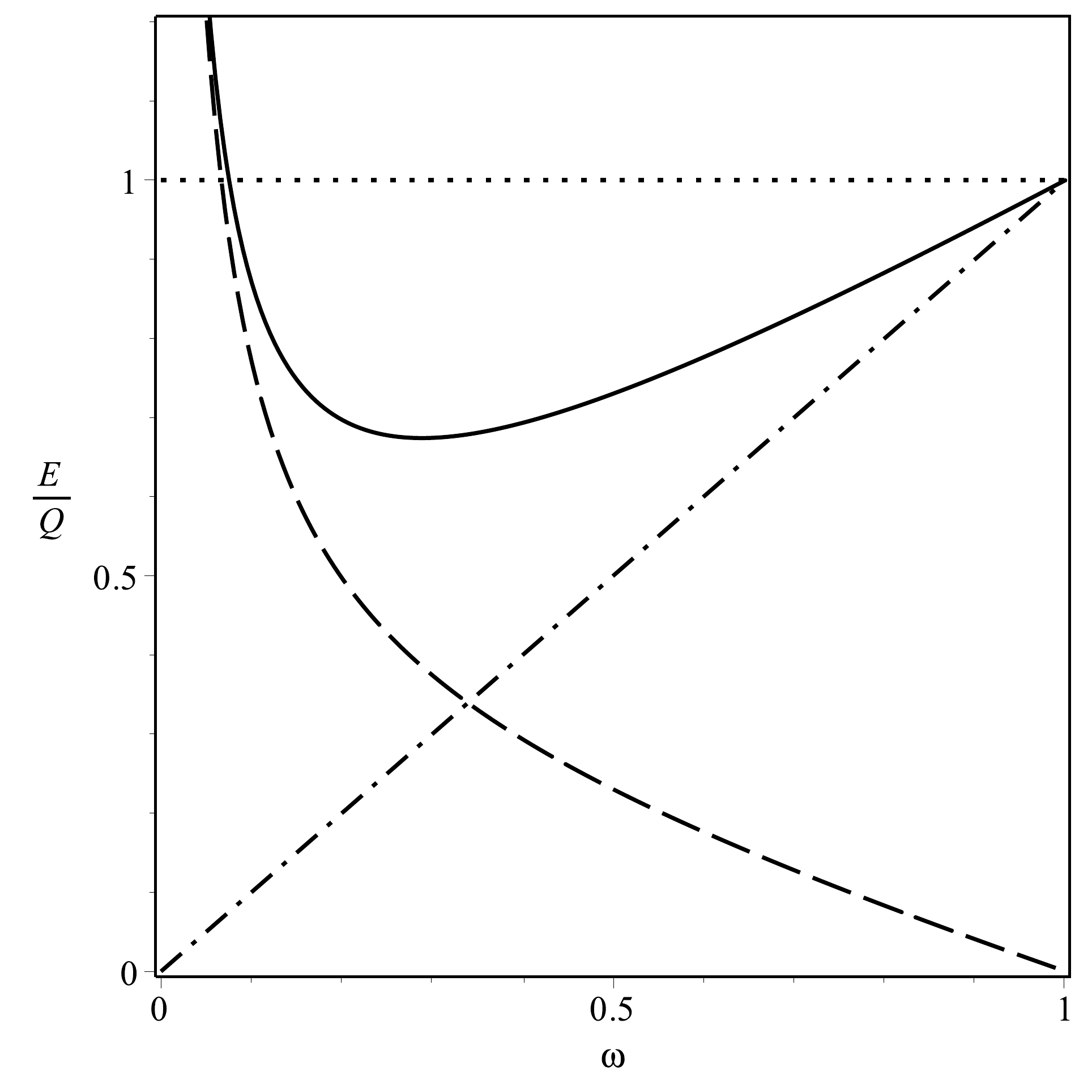}
\includegraphics[width=4.2cm]{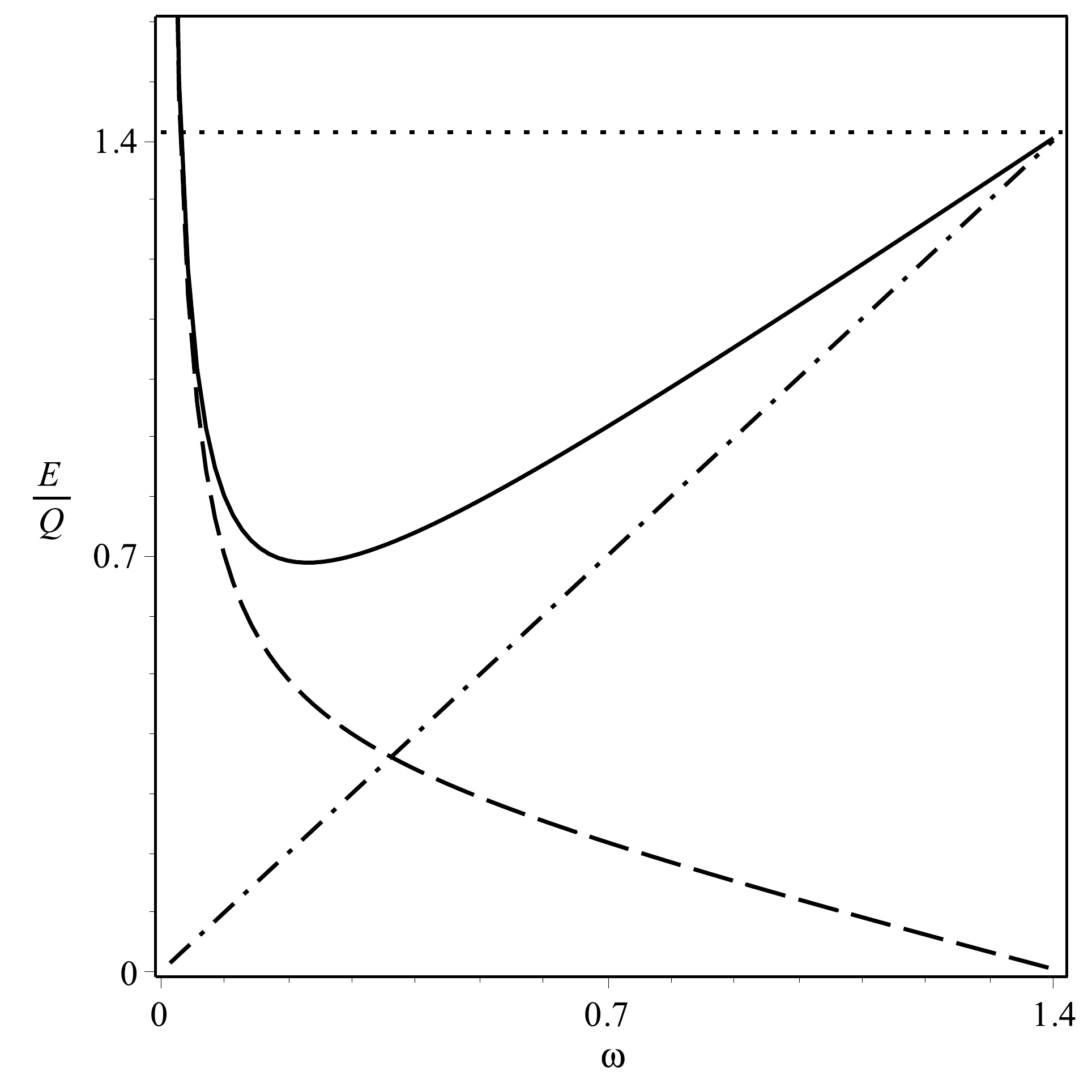}
\includegraphics[width=4.2cm]{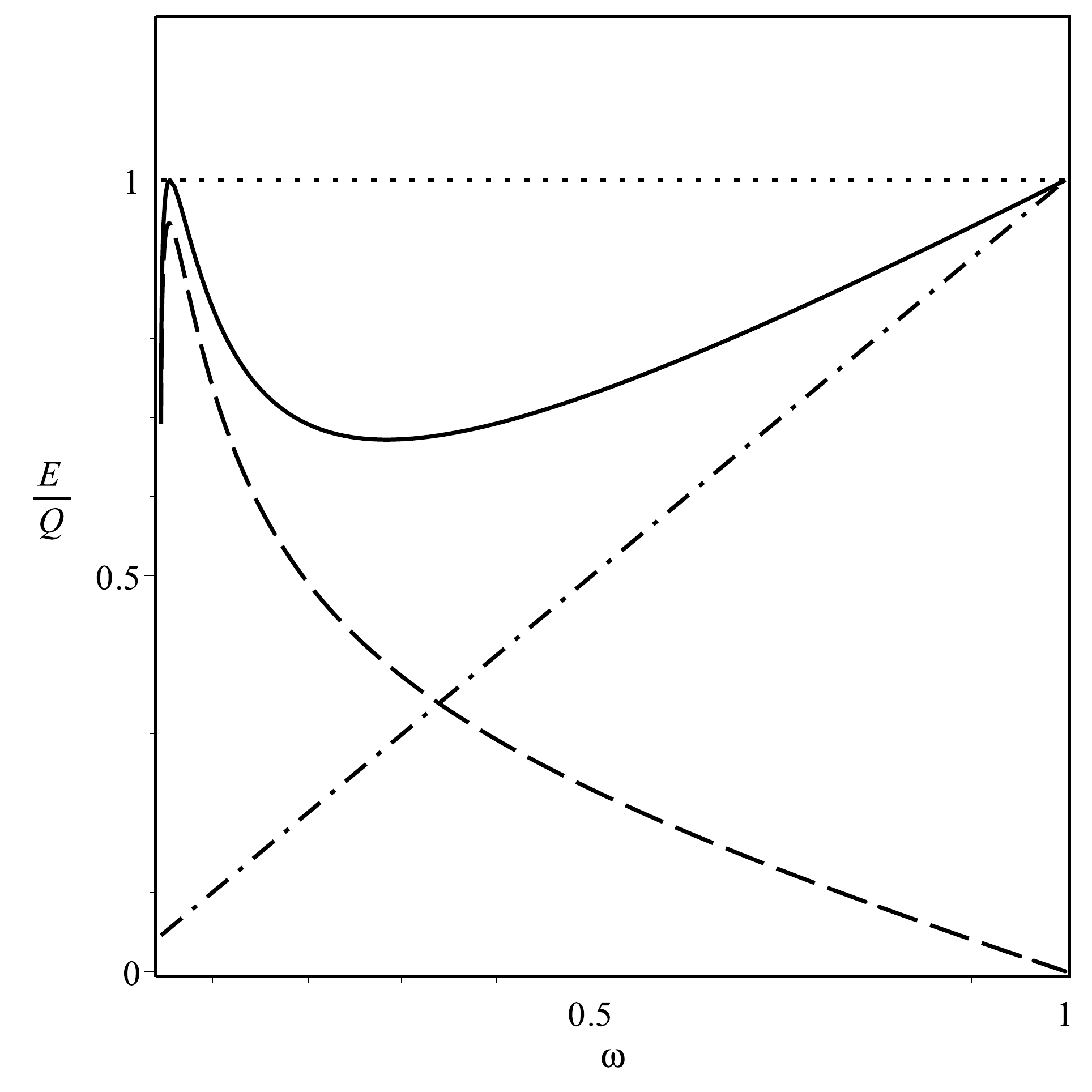}
\includegraphics[width=4.2cm]{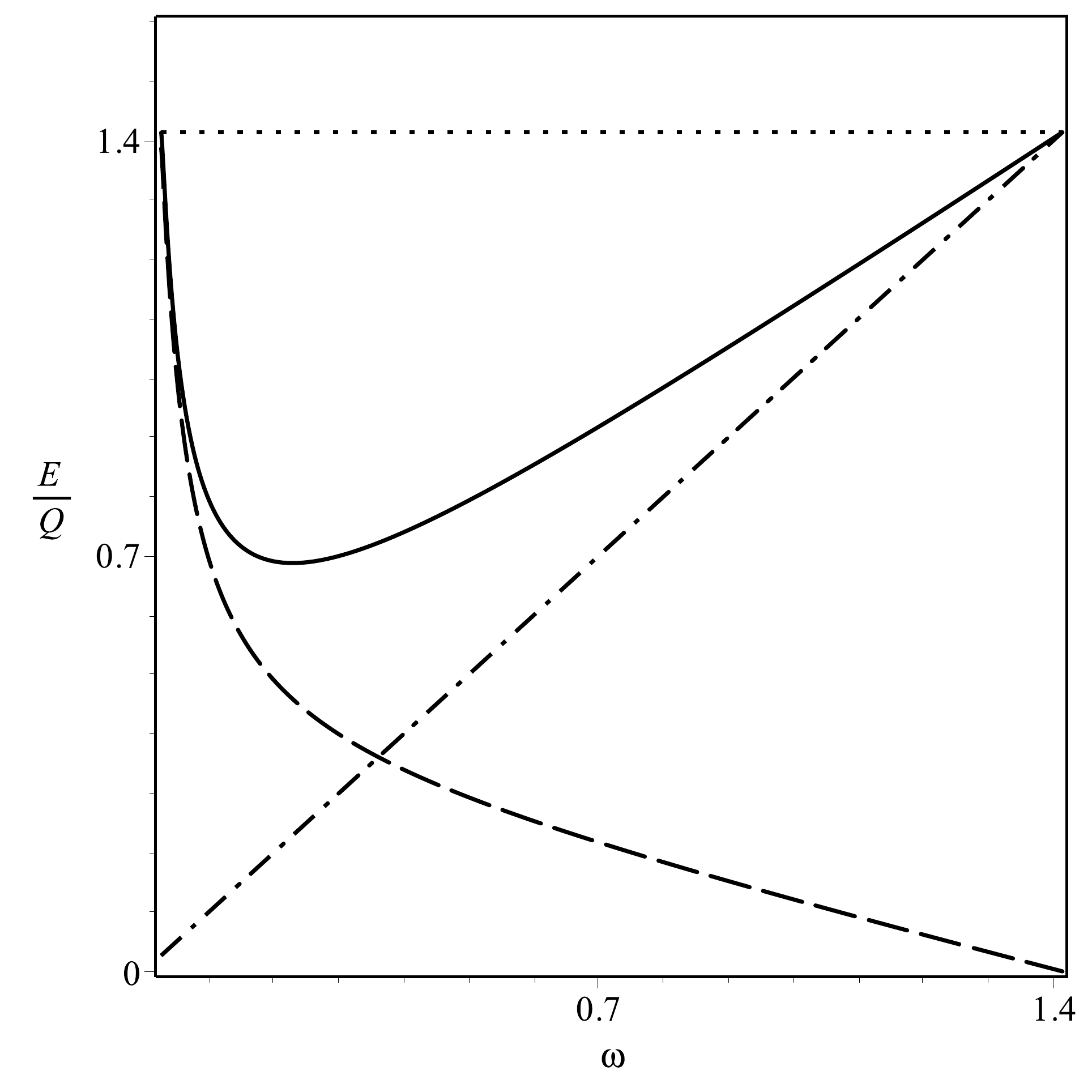}
\includegraphics[width=4.2cm]{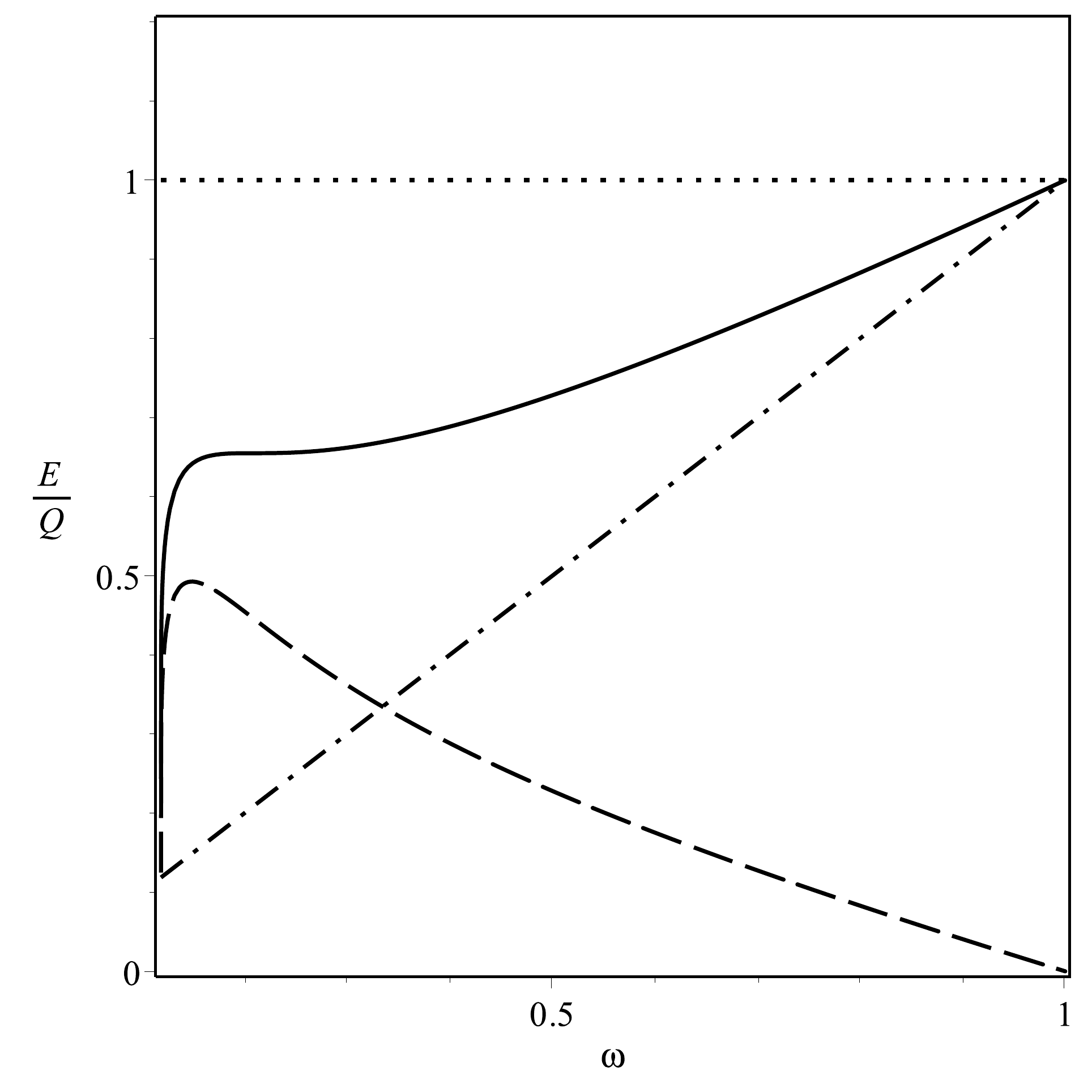}
\includegraphics[width=4.2cm]{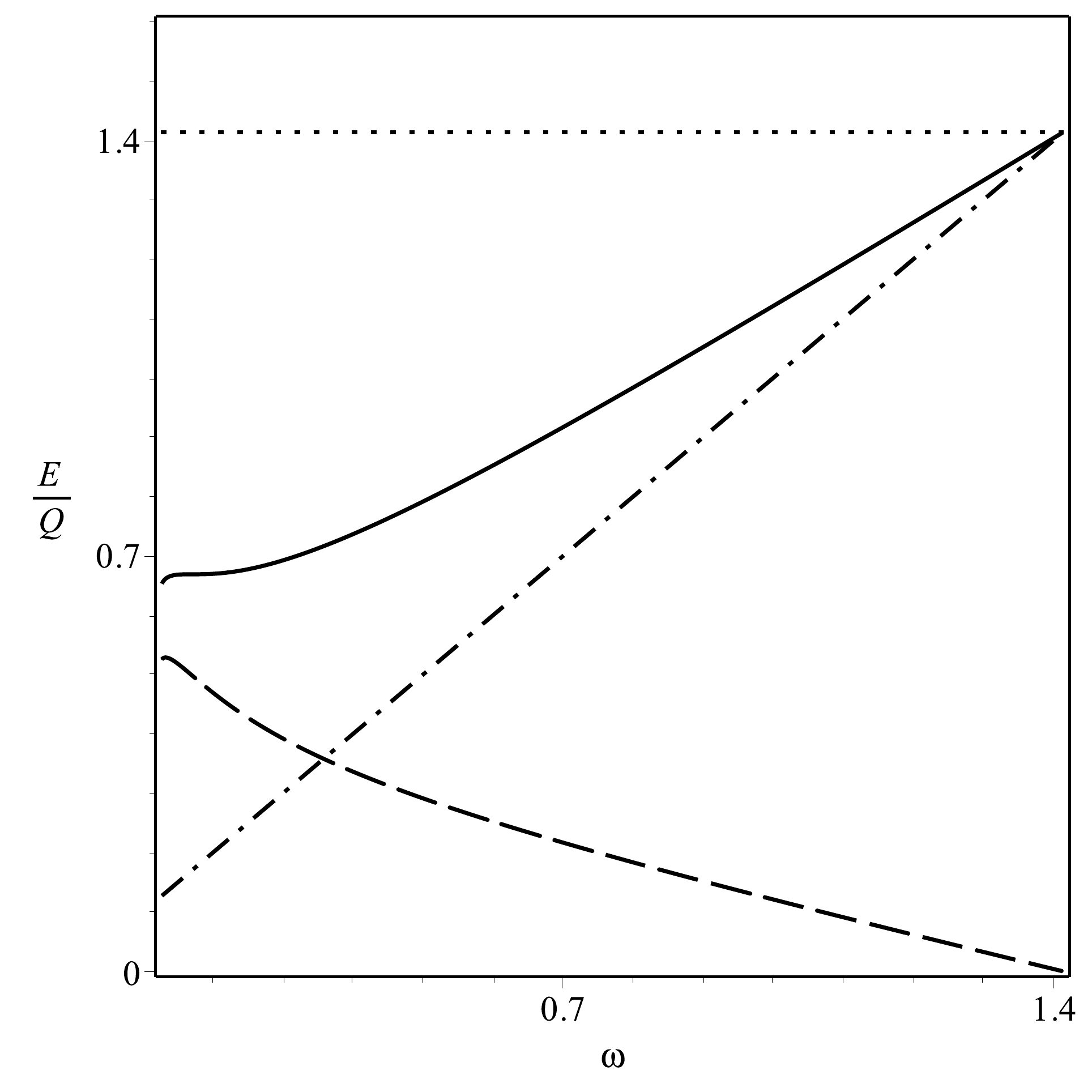}
\includegraphics[width=4.2cm]{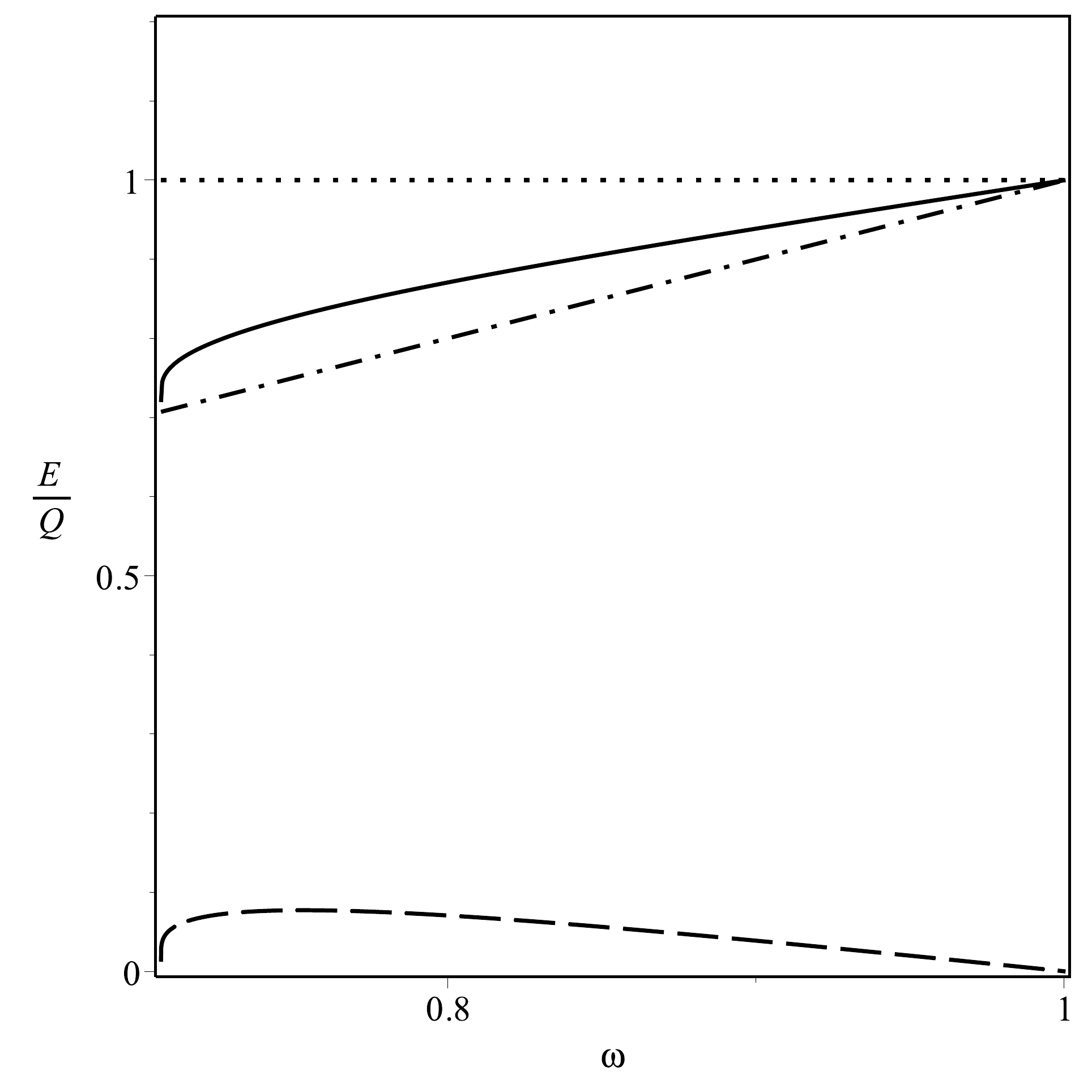}
\includegraphics[width=4.2cm]{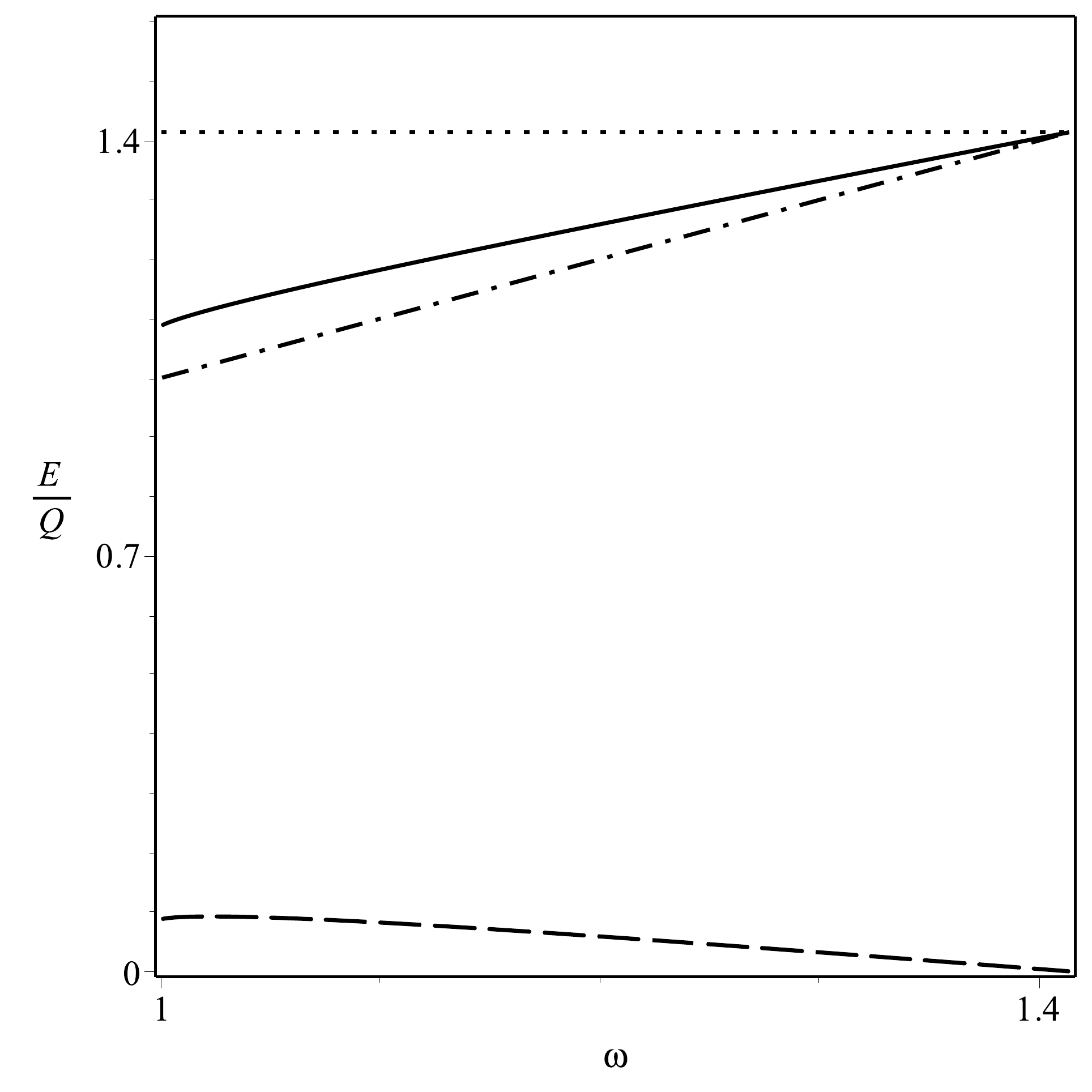}
\caption{The ratio $E/Q$ between the energy and the charge as a function of $\omega$ in the standard (left) and generalized (right) models, for $a=a_0$ (top), $ a_{s1}$ and $a_1$ (middle-top), $a_{s2}$ and $a_2$ (middle-bottom), and for $a_3$ (bottom). The dash-dotted and the dashed lines represent the ratio with the charge dependent $(E_Q)$ and the charge independent $(E_I)$ portions of the energy, and the solid lines stand for the ratio with the total energy. The dotted line represents $\omega_+$.}
\label{eqplot}
\end{figure}

Before ending the work, let us further comment on two distinct behaviors of the new model, the first one related to the field having the form of a plane wave, $\vphi=\alpha e^{ik_\mu x^\mu}$. We suppose that
$\alpha\ll1$, so we can consider the potential as $V\approx |\vphi|^2/2$. In this case the equation of motion \eqref{eomf} with $f=1/(4V)$ provides the dispersion relation $k_\mu k^\mu =\omega_+^2$. Thus, the free particle has energy $E=\omega_+Q$ and the criteria for quantum mechanical stability does not change from the standard case, being $E/Q<\omega_+$.

The second behavior concerns the ratio $E/Q$, which is depicted in Fig.~\ref{eqplot} for the standard and generalized models. We note that, starting from $a=a_0$, there appears an interval of $\omega$ such that $E/Q>\omega_+$, where the solution is unstable. We increase $a$ up to $a_1$. In this case there is a value of $\omega$ in which $E/Q=\omega_+$. Thus, the solution is quantum mechanically stable if $a>a_1$. Here we remark that the above Eqs.~\eqref{qwm} and \eqref{eiwm} are important to find the numerical value of $a_1$.

\section{Conclusions}\label{conclusions}

In this work we investigated the presence of non-topological solutions of the Q-ball type, in models described by a single complex scalar field having global $U(1)$ symmetry. We first reviewed the standard case and then moved on to propose and investigate another model, much more involved than the standard one. Due to the modification included in the Lagrange density, we studied the model under the assumption that the solution depends only on $x$ and $t$, so we worked in $(1,1)$ spacetime dimensions. 

Despite the modification introduced, we have been able to find explicit analytical solution for the new model, which is inspired by the k-field type of kinematical modification. The solution is referred to as a split Q-ball configuration. The presence of the exact solution helped us to obtain exact expressions for the charge and energy and to analyze its quantum mechanical stability, as a function of $\omega$ and of the parameter that controls the potential of the model. As shown analytically, the solution found in the generalized model is similar to the standard solution, but the profile of the charge and energy densities show a splitting around its center, which simulates an internal structure.

The new structure, the analytical non-topological solution of the Q-ball type that we presented in this work appears to be of interest to phenomenologists. Thus, it seems that further investigations are needed, in particular in the more realistic scenario that includes the three dimensional space and the local $U(1)$ symmetry. One expects that the splitting found in the one-dimensional case may contribute to form a hollow ball configuration in the three-dimensional space. We are now dealing with such issues, and hope to report on them in the near future.

\acknowledgements{We would like to thank the Brazilian agency CNPq for partial financial support. DB thanks support from projects 455931/2014-3 and 306614/2014-6, LL thanks support from projects 307111/2013-0 and 447643/2014-2, MAM thanks support from project 140735/2015-1, and RM thanks support from projects 455619/2014-0 and 306826/2015-1.}

\end{document}